\documentclass[english,pre,floats,superscriptaddress,usenames]{revtex4}
\usepackage[T1]{fontenc}
\usepackage[latin9]{inputenc}
\usepackage{amsmath}
\usepackage{amssymb}
\usepackage{graphicx}
\usepackage{esint}

\makeatletter

\@ifundefined{textcolor}{}
{%
 \definecolor{BLACK}{gray}{0}
 \definecolor{WHITE}{gray}{1}
 \definecolor{RED}{rgb}{1,0,0}
 \definecolor{GREEN}{rgb}{0,1,0}
 \definecolor{BLUE}{rgb}{0,0,1}
 \definecolor{CYAN}{cmyk}{1,0,0,0}
 \definecolor{MAGENTA}{cmyk}{0,1,0,0}
 \definecolor{YELLOW}{cmyk}{0,0,1,0}
 }

\@ifundefined{definecolor}{\usepackage{color}}{}

\newcommand{\be}{\begin{equation}} 
\newcommand{\ee}{\end{equation}}
\newcommand{\bea}{\begin{eqnarray}}   
\newcommand{\eea}{\end{eqnarray}}

\newcommand{\vv}{{\bf v}}
\newcommand{\cc}{{\bf c}}
\newcommand{\uu}{{\bf u}}
\newcommand{\rr}{{\bf r}}
\newcommand{\FF}{{\bf F}}
\newcommand{\GG}{{\bf G}}
\newcommand{\bk}{{\bf k}}
\newcommand{\JJ}{{\bf J}}

\newcommand{\CC}{{\bf C}}

\newcommand{\fa}{f^{\alpha}}
\newcommand{\ga}{g^{\alpha}}
\newcommand{\na}{n^{\alpha}}
\newcommand{\nb}{n^{\beta}}
\newcommand{\Pia}{\Pi^{\alpha}}
\newcommand{\uua}{{\bf u}^\alpha}
\newcommand{\uub}{{\bf u}^\beta}
\newcommand{\gab}{g_{\alpha\beta}}
\newcommand{\sab}{{\bf \sigma}_{\alpha\beta}}

\usepackage{babel}

\makeatother

\usepackage{babel}
\begin{document}

\date{\today}

\title{Stabilized Lattice Boltzmann-Enskog method for compressible flows
and its application to one and two-component fluids in nanochannels }

\author{Simone Melchionna}

\address{CNR-IPCF, Consiglio Nazionale delle Ricerche, 
Universit\`a di Roma La Sapienza, P.le A. Moro 2,
00185 Roma, Italy}

\author{Umberto Marini Bettolo Marconi}

\email{umberto.marinibettolo@unicam.it}

\selectlanguage{english}

\address{ Scuola di Scienze e Tecnologie, Universit\`a di Camerino, Via
Madonna delle Carceri, 62032 , Camerino, INFN Perugia and CNISM, Italy}
\begin{abstract}
{A  numerically stable method to solve 
the discretized Boltzmann-Enskog equation describing the behavior of non ideal fluids 
under inhomogeneous conditions is presented.  The algorithm employed uses
a Lagrangian finite-difference scheme for the treatment of the convective
term and a forcing term to account for the molecular repulsion together
with a  Bhatnagar-Gross-Krook  relaxation term. In order to eliminate the spurious currents
induced by the numerical discretization procedure, we use a trapezoidal 
rule for the time integration together with a version of the two-distribution
method of He {\em et al.} (J. Comp. Phys \textbf{152}, 642 (1999)). 
Numerical tests show that, in the case of
one component fluid in the presence of  a spherical potential well,
the proposed method  reduces the  numerical error by several orders of magnitude. 
We conduct another  test by considering the  flow of  a  two component fluid in a channel with a bottleneck and provide information about the density
and velocity field in this structured geometry.}

\end{abstract}
\maketitle

\section{Introduction}

Liquids often appear as homogeneous on a macroscopic scale, but not when observed
on a microscopic scale where they may display density oscillations extending over 
a few molecular diameters.
Equilibrium statistical mechanics theories such as density-functional theory (DFT)  or integral equations  can
deal routinely with the presence of such inhomogeneities in density,
concentration or other kinds of order parameters, and predict 
the ensemble average microscopic profiles and
the associated surface and line tension, while a similar
situation does not occur in non-equilibrium systems \cite{Evans,hansen,wu}.
In this case, the presence of inhomogeneities  often
causes difficulties in the numerical solution of the associated evolution equations.

It is well known that the conventional hydrodynamic description, based on the Navier-Stokes equation,
faces difficulties when fluids are confined within a small volume or when the boundaries of the container have  complicated
shapes with typical lengths of the order of  a few molecular diameters. Such
a picture, while valid on a macroscopic scale, fails to describe very
small systems \cite{NavierStokes,Bruus,Molecularphys2011}.
On the other hand, the kinetic approach based on the distribution
functions formalism and on the Boltzmann equation and its refinements represents a convenient description of both homogeneous
and inhomogeneous systems. Among the existing numerical approaches employed
to solve the Boltzmann equation,
the Lattice Boltzmann (LB) method plays a prominent role
\cite{LBgeneral,LBmicro,Ansumali}. 
It is a discretized
version of the continuous Boltzmann equation and gives good results
in the homogeneous phases \cite{HeLuo,HeLuo2,Abe}. However, the numerical solution
of inhomogeneous systems within the LB scheme is challenging:
as reported by several authors \cite{Wagner,Shan,TOV,Guo1,Pooley,Zhang},
a straightforward application of the LB equation leads to the observation of an 
unphysical effect, the so-called spurious currents, resulting from the discretization procedure.
To cure this pathology, molecular interactions must be handled with care.

In the literature, internal forces are accounted for in
two different ways: a) either by imposing the condition that the equilibrium
distribution gives the desired form of the pressure tensor or b) by
introducing an appropriate forcing term \cite{kikkinides,Greci}.
The forcing term can be chosen in two different manners, either 
proportional to the gradient of the pressure excess over the ideal
gas value or proportional to the product of the density times the gradient
of the excess chemical potential, that is, by using the Gibbs-Duhem condition
in differential form. Actually, the second choice
is consistent with microscopic theories, such as DFT \cite{Evans}, 
where the equilibrium condition is given by requiring
that  the gradient of the local chemical potential is locally balanced by the external forces.

In the present paper, we discuss a LBE algorithm based on the Boltzmann-Enskog transport
equation \cite{vanbeijeren,Santos,Santos2,Lutsko,Anero}.
 The approach is particularly convenient when the packing effects
 are relevant, that is from moderate to high  fluid densities.
 A straightforward application of the LBE algorithm leads to numerical instabilities
 so  we introduce a numerical scheme that employs a trapezoidal time discretization
plus an extension of a procedure, originally proposed by He {\em et al.} \cite{He}, 
that uses two distribution functions, instead of one, to reduce the spurious currents 
phenomenon.
 In this scheme, one distribution function tracks
the local density profile while the other tracks  the local momentum density. 
The standard phase space distribution function $f(\rr,\vv,t)$ is evolved concurrently with
an auxiliary distribution function, named $g(\rr,\vv,t)$, whose zeroth
velocity-moment is the hydrodynamic pressure and its first moment
is identical to the corresponding moment of $f(\rr,\vv,t)$. According to previous authors,
the reason for the increased stability of the double
distribution method stems from
the fact that the forcing term in the g-equation is multiplied
by the difference between the local and the global Maxwellian thus
reducing its importance with respect to the original f-equation, where
the forcing term is multiplied by the local Maxwellian. The method
was later extended and generalized by T. Lee and coworkers 
\cite{LeeLin,LeeFischer,Lee}.

The main difference between our approach and the previous ones, besides
the bottom-up microscopic modeling of the fluid proposed in earlier work 
 \cite{Melchionna2008,Lausanne2010,JCP2007},  consists in the
choice of the  function employed to define the g-distribution
function. As we shall see, with the present choice it is straightforward
to generalize the method to multicomponent fluids, while in the
original formulation such a generalization is not straightforward.
In this way, our method leads naturally to a
form of the forcing term similar to that in the Gibbs-Duhem route. This
strategy can also be generalized to multicomponent fluids, whereas
the pressure route cannot.

The paper is organized as follows: in Sec.
\ref{tandem} we present the evolution
equation for the one particle distribution function $f$
and for the auxiliary distribution $g$ both for the simple fluid and for the fluid mixture.
In Sec. \ref{numericalsolution}
we discuss the discretization procedure
In Sec. \ref{results} we present numerical tests 
of the proposed method.
  Finally in Sec. \ref{conclusions} we present our conclusions
and perspectives. 

\section{Equations for the double distribution functions}
\label{tandem}

We start the discussion with the set of Boltzmann-Enskog equations characterizing a mixture of $M$ species,
labelled with an upper index $\alpha=1,M$. The evolution equation for a particular distribution function $\fa(\rr,\vv,t)$ 
can be written as: 
\bea      
\frac{D }{Dt} \fa(\rr,\vv,t)=
-\frac{\FF^{\alpha}(\rr)}{m}\cdot
\frac{\partial}{\partial \vv} \fa(\rr,\vv,t)
+ \sum_\beta J^{\alpha\beta}(\rr,\vv,t)
\nonumber\\
\label{uno}
\eea
 where the material derivative is given by: 
\begin{equation}
\frac{D}{Dt}=\frac{\partial}{\partial t}+\vv\cdot\nabla.
\end{equation}
and 
$\FF^{\alpha}(\rr)$ is an external velocity independent force field acting on component $\alpha$ and
$J^{\alpha\beta}$,
represents the effect on the single particle distribution function of the
interactions among the fluid particles of  type $\alpha$ and $\beta$.

Using a separation of the interaction term into  
a kinetic rapidly varying part and an hydrodynamic part 
originally introduced by Santos et al. \cite{Santos,Santos2} and
extended to 
mixtures later \cite{Melchionna2009,JCP2011}
we rewrite \eqref{uno} as: 
\begin{equation}
\frac{D \fa}{Dt}=-\omega(\fa-\fa_{eq})+S^\alpha_f(\rr,\vv,t)
\label{boltzoriginal}
\end{equation}

The first term in the r.h.s. of eq. \eqref{boltzoriginal} is 
a Bhatnagar-Gross-Krook (BGK) relaxation term \cite{BGK}, 
$\omega$ an inverse relaxation time, and $S^\alpha_f$
is a source term due to external forcing and molecular
interactions. According to  \cite{JCP2011b} it can be written as
\begin{equation}
S^\alpha_f(\rr,\vv,t)=-\frac{\FF^\alpha(\rr)}{m}\frac{\partial}{\partial\vv}\fa+\beta(\vv-\uu)\cdot\CC^\alpha(\rr,t)\Gamma_{u}(\rr,\vv,t)
\label{boltzoriginal-1}
\end{equation}
 with $\beta=1/k_{B}T$, $T$ is the temperature and $k_{B}$ the Boltzmann
constant. In addition, $\Gamma_{u}$ is a Maxwellian velocity distribution
whose mean velocity is the local  fluid velocity $\uu(\rr,t)$: 
\begin{equation}
\Gamma_{u}(\rr,\vv,t)=\Bigl(\frac{1}{2\pi v_{T}^{2}}\Bigl)^{3/2} e^{-(\vv-\uu(\rr,t))^{2}/2v_{T}^{2}}
\end{equation}
 where $mv_{T}^{2}=k_{B}T$ for particles of common mass $m$, and
 \be
f^{\alpha}_{eq}(\rr,\vv,t)=n^{\alpha}(\rr,\vv,t) \Bigl\{1+
\beta [(\uua(\rr,t)-\uu(\rr,t))\cdot(\vv-\uu(\rr,t))]\Bigl\}
\Gamma_u(\rr,\vv,t) 
\label{prefactor}
\ee
with $\uua$ the average velocity of the component $\alpha$.
The term $\CC^\alpha$ is a collisional kernel describing the change of $\fa$ due to the interactions. 

We first rewrite \eqref{boltzoriginal} in a form that  is equivalent
up to terms of third order in the Hermite expansion
\begin{equation}
S^\alpha_f=\beta \left(\CC^\alpha+\na\frac{\FF^\alpha}{m}\right)\cdot(\vv-\uu)\Gamma_{u}.
\label{bolt2}
\end{equation}

From the phase space distribution function $\fa(\rr,\vv,t)$
we can compute the particle partial density,
\be
\na(\rr,t)=\int  d\vv \fa(\rr,\vv,t)
\ee 
and the momentum current carried by particles of type $\alpha$,
\be
\na(\rr,t)\uua(\rr,t)=\int  d\vv \fa(\rr,\vv,t)\vv
\ee 
which from eq. \eqref{boltzoriginal} satisfies the continuity equation
\begin{equation}
\frac{\partial \na}{\partial t}+\nabla\cdot(\na\uua)=0\label{continuity}.
\end{equation}
 The average fluid velocity is obtained from 
 $$\uu=\frac{\sum_\alpha\na\uua}{n}$$ with the global density given by $$n=\sum_\alpha \na.$$
  
The numerical solution of eq. \eqref{boltzoriginal} is plagued by
numerical instabilities as reported in ref. \cite{He}, because the term $S^\alpha_f$ featured in
the r.h.s. is quite large in the interfacial regions, since the main contribution to $\CC^\alpha$,
which is proportional to the gradient of the local chemical potential, varies rapidly.
Alternatively, following the seminal idea put forward by He and coworkers in ref. \cite{He} 
and pursued by Lee and coworkers \cite{Lee}  to stabilize the numerical
solution the one component version of eq. \eqref{boltzoriginal}, it is possible to employ an
auxiliary distribution function, named $\ga(\rr,\vv,t)$ such that the role of the
forcing term  featured in its evolution equation is effectively reduced.
Such an heuristic recipe stabilizes the numerical
solution  by
\textquotedbl{}decoupling\textquotedbl{} the density and the momentum
equations. In the present treatment, we will handle the stabilizing
terms in an effective way, without relying on any heuristics.
Let us introduce the auxiliary distribution 
\be
\ga(\rr,\vv,t)=\fa(\rr,\vv,t)+(\Pia(\rr,t)-\na(\rr,t))\Gamma_0
\label{definition} 
\ee
 where $\Pia(\rr,t)$
  is a function of the partial densities 
, to be determined in the following, and $\Gamma_{0}$ indicates the velocity distribution at  global equilibrium,
that is, the Maxwellian corresponding to $\uu=0$.
One assumes that
the function $\Pia$ depends from its argument through $\{\na(\rr,t)\}$.
From the definition \eqref{definition} one can see that
 $\ga$ differs from $\fa$ w.r.t. the zeroth moment
\be
\int d\vv \ga(\rr,\vv,t)=\Pia(\rr,t) ,
\ee
but shares the same first moment 
\be
\int d\vv \ga(\rr,\vv,t)\vv =\na(\rr,t)\uua(\rr,t).
\ee

By using eqs.\eqref{definition} and \eqref{boltzoriginal}, 
the evolution equation for $\ga(\rr,\vv,t)$ reads
\begin{equation}
\frac{D\ga}{Dt}=\frac{D \fa}{Dt}+\frac{D}{Dt}(\Pia-\na)\Gamma_{0} .
\label{gtime}
\end{equation}
\be
 \frac{D}{D t}(\Pia(\rr,t)-\na(\rr,t)) = (\vv-\uu)\cdot\nabla (\Pia-\na) 
 -\sum_\beta\nb (\frac{d \Pia}{d\nb }-\delta_{\alpha\beta})\nabla \cdot \uub 
 -\sum_\beta(\frac{d \Pia}{d\nb }-\delta_{\alpha\beta})(\uu^\beta-\uu)\cdot\nabla \nb    \ee 
  one obtains the evolution equation  \eqref{gtime} for $\ga(\rr,\vv,t)$ as
\be
\frac{D \ga}{D t}=-\omega (\ga-\ga_{eq})+ S^\alpha_g
\label{eqg}
 \ee
 with
 \bea
 &&
 S^\alpha_g(\rr,t)=\beta\Bigl( {\bf C}^{\alpha} + \na\frac{ \FF^{\alpha}}{m}\Bigl)\cdot(\vv-\uu) (\Gamma_u-\Gamma_0)
\nonumber\\
&&+\Bigl[ \nabla \Pia-\nabla \na+\beta {\bf C}^{\alpha} +  \na \frac{\FF^{\alpha}}{m}\Bigl] \cdot (\vv-\uu) \Gamma_0 
-\sum_\beta \nb (\frac{d \Pia}{d\nb }-\delta_{\alpha\beta})(\nabla \cdot(\nb \uub )-\uu\cdot\nabla \nb) \Gamma_0 
 \label{palfa}
 \eea 
 and
 \be
 \ga_{eq}(\rr,\vv,t) =\na(\rr,t)  \Bigl (1+
\beta m(\uua(\rr,t)-\uu(\rr,t))\cdot(\vv-\uu(\rr,t))\Bigl) \Gamma_u+(\Pia(\rr,t)-\na(\rr,t))\Gamma_0 .
 \ee 
It can be checked that
the evolution equation for  $\ga$
or the one for $\fa$ lead to the same balance equation for $\uua$. 
In practice, in the numerical work  we shall use the $\fa$ equation to determine the density
 $\na$ and track the formation of interfaces,
and the $\ga$ equation to determine the velocity field $\uua$.


 The main motivation behind the transformation from $\fa$ to $\ga$
 is that  the effect of the forcing  term, $S^\alpha_g$, featuring in \eqref{eqg} can be rendered smaller
 than the corresponding effect due to    the forcing  term, $S^\alpha_f$, in  the original equation  \eqref{boltzoriginal} for $\fa$ 
 by an appropriate choice of the function $\Pia(\rr,t)$.
 In fact, the first  term in $S^\alpha_g$ is of order $(\vv-\uu)^{2}$ because it contains the product of $(\vv-\uu)(\Gamma_u-\Gamma_0)$ ,
 whereas the second term can be rendered small using the arbitrariness of the function
$\Pia$ is. As far as the last term is concerned
 we shall verify that the  last term in $S^\alpha_g$ is 
actually small in our  numerical simulation.
One expects that a weaker forcing term helps the stability of the numerical solution.
In the one component case  He and coworkers
suggested to replace  $\beta^{-1}\Pia$ by the  thermodynamic pressure $p_t$.
In order to see that we use the explicit representation of the function $\CC^\alpha$, which represents
the effect of the molecular interactions in the model studied.

We first separate the effective field $\CC^\alpha$ into three separate contributions,
the separation being quite generic and not determined by the particular model used:
\be
{\bf C}^{\alpha}(\rr,t)= \CC^{\alpha,mf}(\rr,t)+\CC^{\alpha,drag}(\rr,t)+\CC^{\alpha,visc}(\rr,t) .
\label{splitforce}
\ee
The first term can be written as:
\be
\CC^{\alpha,mf}= -\na(\rr,t)\nabla\mu_{int}^{\alpha}(\rr,t).
\label{potchimico2}
\ee
where $ \mu_{int}^\alpha$ is the non ideal part of the chemical
potential of the $\alpha$ component.
For density profiles smooth enough we can write 
\be
\CC^{\alpha,drag}(\rr,t) \simeq
-\gamma \na(\rr,t) \sum_\beta (\uua(\rr,t)-\uub(\rr,t)) 
\label{dragforce2}
\ee
and for the viscous part
\bea
C_i^{\alpha,visc}(\rr,t) 
\approx -\na(\rr,t)\sum_\beta\Bigl( \eta^{\alpha\beta} \nabla^2 u^\beta_i(\rr,t)+(\eta^{\alpha\beta}_b+\frac{1}{3}\eta^{\alpha\beta})\nabla_i(\nabla\cdot \uub)\Bigl)
\label{viscousforce2}
\eea
The coefficients $\gamma$ and $\eta$ depend on the specific model. In appendix A, we report their explicit representation for 
a system of hard-spheres with attractive interactions. 

In the case of a one component fluid it is straightforward to derive the equation for the $g$ distribution which closely
resembles the equation derived by He and coworkers. After dropping the unnecessary index $\alpha$ one has:
\begin{eqnarray}
\frac{Dg}{Dt}  =-\omega(g-g_{eq}) + S_g
\label{eq:g-evolution}
\end{eqnarray}

\begin{eqnarray}
S_g  = \beta (\CC + n \frac{\FF}{m} )\cdot(\vv-\uu)(\Gamma_{u}-\Gamma_{0})+\left[\nabla \Pi-\nabla n+\beta \CC
 +n \frac{\beta\FF}{m}  \right]\cdot(\vv-\uu)\Gamma_{0}-n(\frac{d\Pi}{dn}-1)(\nabla\cdot\uu)\Gamma_{0} . \nonumber\\
\label{pg}
\end{eqnarray}
Using \eqref{potchimico2} and neglecting the non equilibrium contributions to $\CC$ we have:
\be
-\nabla n+\beta \CC=-\beta n\nabla\mu
\ee
where  $\mu$ is the total chemical potential. Finally
with the help of the Gibbs-Duhem relation we introduce the thermodynamic pressure :
\be
\nabla p_t= n\nabla \mu .
\ee
Hence, requiring the vanishing of the square brackets in \eqref{pg} 
is equivalent to the condition:
 \be
 \beta\nabla \Pi =\nabla p_t- n\frac{\FF}{m} .
 \ee
In other words choosing $\beta\Pi$ to be the thermodynamic potential augmented by the contribution due to the external field
makes the second term of \eqref{pg} to vanish. From the physical point of view such a condition is a consequence 
of the hydrostatic equilibrium condition \cite{Evans}.

Unfortunately, in the multi-component fluid the identification of $\Pia$ with the pressure is not possible.
 The reason is that in this case the distribution functions one needs a $\Pia$ function for each component,
 whereas one can find only one pressure, through the 
 Gibbs-Duhem relation
\be
\nabla p_t= \sum_\alpha\na\nabla \mu^\alpha .
\ee
Moreover,
by using the pressure route it is very difficult to obtain a satisfactory numerical solution 
in the general case, as in presence of confining walls,
spontaneous layering mechanisms, or free interfaces.
Alternatively, we choose the unknown function
$\Pia$ as the \textquotedbl{}potential function\textquotedbl{} associated
with the vector field $\nabla \na-\beta \CC^{\alpha}$, in such
a way as to cancel this term from eq. \eqref{palfa}.
More precisely, the function $\Pia$ is chosen to be:
 \begin{equation}
\Pia(\rr,t)=\na(\rr,t)-\beta \int^{\rr}d\rr' \Bigl(\CC^{\alpha,mf}(\rr',t) +\na(\rr',t)\frac{\FF^{\alpha}(\rr')}{m}\Bigl)
\label{quasipressure}
\end{equation}
Therefore, being $\CC^{\alpha,mf}$ a functional of density, $\Pia$ is chosen to 
be a non-local function of density, in stark contrast with previous proposed approaches that
are based on a local compensating pressure term \cite{He,Lee,Karni,Karni2}.
Eq. (\ref{quasipressure}) also provides the operational route to our approach. In fact,
the integral is evaluated numerically
using trapezoidal spatial integration, which provides a satisfactory numerical
solution in terms of accuracy. It should be noticed that, being an
integral over a vector field, the integration depends on the
origin and the specific path of the integral. However, this aspect
is not dangerous for systems where a symmetry point can be found.
In addition, the integration constant never appears in the evolution
equation and thus does not need to be determined.
Although eq. \eqref{eqg} looks more complicated
than the original one, it behaves better in numerical terms and
gives rise to smaller interfacial currents, as shown in the sequel.


\section{Numerical solution }
\label{numericalsolution}
We illustrate the numerical solution of the proposed method by considering explicitly 
the one component case, while the multicomponent case can be easily deduced.
Let us consider again the integration of the generic evolution equation
\begin{equation}
\frac{Df}{Dt}=\Omega(f,M)(\rr,\vv,t)\label{bolt1}
\end{equation}
 where the unspecified kernel $\Omega$ contains both the collisional term,
the BGK term and the external force ${\bf F}\cdot\partial_{\vv}f$.
{As customary in the derivation of the Lattice Boltzmann method, the distribution
function is first projected on an finite Hermite basis set to handle the dependence
on velocity \cite{shanyuanchen,Moroni}.}
By taking eq. \eqref{bolt1} as our reference equation, the r.h.s. depends
on $f$ but also on its moments $M=\{M_{p}\}$, with 
\begin{equation}
M_{p}(\rr,t)=<f|{\cal H}_{p}>
\end{equation}
 where ${\cal H}_{p}$ is the p-th Hermite polynomial, and 
\begin{equation}
<A|{\cal H}_{p}>\equiv\int d\vv A(\rr,\vv,t){\cal H}_{p}(\vv)
\end{equation}
 expresses the Hermite scalar product.

In order to discretize eq. (\ref{bolt1}), we start by considering
the following truncated Hermite expansion 
\begin{equation}
\bar{f}(\rr,\vv,t)=\Gamma_{0}(v)\sum_{p=0}^{K}\frac{1}{v_{T}^{2p}2p!}M_{p}(\rr,t){\cal H}_{p}(\vv)
\end{equation}
 where $K$ is the order of truncation of the Hermite expansion and $M_{p}=<f|{\cal H}_{p}>=<\bar{f}|{\cal H}_{p}>$.
In fact, from the definitions, it follows that the original and the
truncated forms of the singlet distribution share the same moments
up to $p\leq K$. By the same token, we consider the expansion of
the collisional kernel, 
\begin{equation}
\bar{\Omega}(\rr,\vv,t)=\Gamma_{0}(v)\sum_{p=0}^{K}\frac{1}{v_{T}^{2p}p!}O_{p}(\rr,t){\cal H}^{(p)}(\vv)
\end{equation}
 with $O_{p}=<\Omega|{\cal H}^{(p)}>$. As for the distribution function,
$\Omega$ has moments $O=\{O_{p}\}$ shared by the full and truncated
representations of the kernel $\Omega$.

The LBM is based on replacing the Hermite scalar products by Gauss-Hermite
quadratures to evaluate its moments, 
\begin{equation}
M_{p}=<\bar{f}|{\cal H}_{p}>=\sum_{p=0}^{G}f_{p}{\cal H}^{(p)}(\cc_{p})
\end{equation}
 where the vectors ${\bf c}_{p}$ are a set of quadratures nodes,
$w_{p}$ are the associated weights, and $G$ is the order of the quadratures.

The operational version of the LBM scheme is provided by the following
quantities $f_{p}(\rr,t)=w_{p}\bar{f}(\rr,{\bf c}_{p},t)/\Gamma_{0}(c_{p})$
and $\Omega_{p}(\rr,t)=w_{p}\overline{\Omega}(\rr,{\bf c}_{p},t)/\Gamma_{0}(c_{p})$.
From these transformations, the evolution equation of the new representation
reads 
\begin{equation}
\frac{\partial }{\partial t} f_p(\vv,t)+\cc_p\cdot\nabla f_p(\vv,t)
=\Omega_{p}(f_{p},M)(\rr,t)\label{eq:bolt2-discrete}
\end{equation}
 where we have rewritten the  streaming term $\vv\cdot\nabla f$  in its Hermite form.
The exact time evolution of the populations over a timestep $h$ then
reads 
\begin{equation}
f_{p}(\rr+\cc_{p}h,t+h)=f_{p}(\rr,t)+\int_{t}^{t+h}ds\Omega_{p}(f_{p},M)(\rr,s)
\end{equation}
 On the other hand, a second-order accurate $O(h^{2})$ numerical
integration can be obtained via the trapezoidal rule \cite{rotenbergmoroni}, 
\begin{equation}
\int_{t}^{t+h}ds\Omega_{p}(f_{p},M)(\rr,s)=\frac{h}{2}\left[\Omega_{p}(f_{p},M)(\rr+\cc_{p}h,t+h)+\Omega_{p}(f_{p},M)(\rr,t)\right]+{\cal O}(h^{3})\equiv\frac{h}{2}\left(\Omega_{p}^{t+h}+\Omega_{p}^{t}\right)+{\cal O}(h^{3})\label{eq:trapezium}
\end{equation}
 where $\Omega_{p}^{t}\equiv\Omega(f_{p},M)(\rr,t)$. 

Eq. \eqref{eq:trapezium} is apparently implicit. However, the scheme
can be rendered explicit by using the following exact mapping to transform
the original populations into the new set 
\begin{equation}
\tilde{f}_{p}=f_{p}-\frac{\Omega_{p}(f_{p},M)}{2}h\label{eq:invertpops}
\end{equation}

The moments in the $\tilde{f}$ representation, collectively called
$\tilde{M}_{p}=\{<\tilde{f}|{\cal H}_{p}>\}$, are related to those
in the $f$ representation by 
\begin{equation}
\tilde{M}_{p}=M_{p}-\frac{O_{p}}{2}h\label{eq:invertmoments}
\end{equation}
 In many circumstances, both relations eqs. \eqref{eq:invertpops}-\eqref{eq:invertmoments}
are invertible, that is, we can obtain explicitly the populations
$f_{p}$ as a function of $\tilde{f}_{p}$ and the moments $M$ as
a function of $\tilde{M}$. This is the case, for example, for BGK
or Fokker-Planck kernels \cite{guo}, and in presence
of external forces.

Finally, the temporal evolution for the populations $\tilde f_{p}$ if
given by the following updating scheme 
\begin{equation}
\tilde{f}_{p}(\rr+\cc_{p}h,t+h)=\tilde{f}_{p}(\rr,t)+\Omega_{p}(f_{p},M)(\rr,t)h\label{eq:newevolution}
\end{equation}
 that provides the way to integrate the equation via the trapezoidal
route. It is practical to work in the $\tilde{f}$ representation,
and substitute the quantities $f_{p}$ and $M$ in the collisional
kernel featuring in the r.h.s. of eq. \eqref{eq:newevolution}.

For the collisional kernel $C$ appearing in eq. (\ref{boltzoriginal}),
however, the relation is non-invertible, since $C$ is a functional
of the hydrodynamic moments. Yet, by decomposing $O(M)=O^{res}(M)+{\bf C}[M]$,
where $O^{res}(M)$ is the residual part of the collisional moments,
being a function (rather than a functional) of the hydrodynamic moments,
a workable algorithm is obtained via the scheme 
\begin{equation}
\tilde{M}_{p}=M_{p}-\frac{O_{p}^{res}(M)}{2}h-\frac{C_{p}[\tilde{M}]}{2}h\label{eq:relatemoments}
\end{equation}
 so that the original moments $M$ are expressed as functionals of
$\tilde{M}$, and substituted in eq. \eqref{eq:newevolution}. It
is straightforward to show that, for $C=0$, the BGK and external
forcing components give rise to the second-accurate integration method introduced
by Guo et al. \cite{guo}.

\section{Results}
\label{results}

In the following, we will analyze an ideal fluid with both the Euler
integration, referred to as EU, and trapezoidal integration, referred
to as TR. Subsequently, we will consider the hard-sphere system and compare
the simulations obtained via the single distribution method (without
the auxiliary distribution), named SD, and the double distribution
method, named DD. By distinguishing the case of Euler integration
from the trapezoidal integration, we have four combinations, for example,
the double distribution method with the trapezoidal rule will be named
DD-TR, and analogously for the other combinations, such as SD-EU,
SD-TR and DD-EU. 


\subsection{Ideal fluid in a potential well}

To illustrate the numerical capabilities of the LBM, let us first
consider an ideal fluid (by setting the collisional kernel $C=0$)
in the presence of an external central potential, expressed as 
\begin{equation}
U^{ext}=\left\{ \begin{array}{c}
-\epsilon(1+\cos(\frac{\pi r}{\xi}))\\
0
\end{array}\right.\;\begin{array}{c}
\mbox{if}\; r<\xi\\
\mbox{else}
\end{array}\label{eq:extpotential}
\end{equation}
where the potential depth is taken to be $\epsilon=0.3\times k_{B}T$
and the well size $\xi$ is varied in order to compare the standard
Euler versus the trapezoidal integration rules. The external force
is expressed as $\FF^{ext}=-\nabla U^{ext}$ acting on particles
of unit mass. At global equilibrium, the density should be distributed
as $n^{eq}_{0}\exp(-U^{ext}/v_{T}^{2})$, with $n_{0}=\frac{1}{V}\int_{V}d\rr n(\rr)$,
and the current should be zero everywhere.

By applying the trapezoidal rule, the populations $\tilde{f}_{p}$
are updated in time and, at every time step the density and current
are computed as $\tilde{n}=\sum_{p}\tilde{f}_{p}$ and $\tilde{\JJ}=\sum_{p}\cc_{p}\tilde{f}_{p}$.
In the $f$ representation, the hydrodynamic moments that contain
the second-order accuracy in space and time are computed by reversing
equation \eqref{eq:relatemoments}, so that $n=\sum_{p}f_{p}=\tilde{n}$
and $\JJ=\sum_{p}\cc_{p}f_{p}=\tilde{\JJ}+\FF^{ext}\frac{h}{2}$.

Finally, the expression of the external forces up to second
Hermite order, reads
\begin{equation}
{\cal F}_{p}^{ext}=w_{p}\left[\FF^{ext}\cdot{\cal H}_{p}^{(1)}+2\FF^{ext}\uu:{\cal H}_{p}^{(2)}\right]
\end{equation}
 where ${\cal H}_{p}^{(1)}={\cal H}^{(1)}(\cc_{p})=\frac{\cc_{p}}{v_{T}^{2}}$
and ${\cal H}_{p}^{(2)}={\cal H}^{(2)}(\cc_{p})=\frac{\cc_{p}\cc_{p}-v_{T}^{2}{\bf I}}{2v_{T}^{4}}$,
being a vector and a tensor of rank two, respectively, and ${\bf I}$ is the unit tensor.

By using eq. \eqref{eq:newevolution}, it follows that $f_{p}=(1+\frac{\omega h}{2})^{-1}\times[\tilde{f}_{p}+\frac{\omega h}{2}f_{p}^{eq}+{\cal F}_{p}]$
and the populations are updated to the following post-collisional
term 
\begin{equation}
\tilde{f}_{p}^{*}=\tilde{f}_{p}+\frac{\omega h}{1+\omega h/2}\left[f_{p}^{eq}(n,\uu)-\tilde{f}_{p}\right]+\frac{h}{1+\omega h/2}{\cal F}_{p}^{ext}
\end{equation}
 to be contrasted with the standard Euler integration, reading
\begin{equation}
f_{p}^{*}=f_{p}+\omega h\left[f_{p}^{eq}(n,\uu)-f_{p}\right]+h{\cal F}_{p}^{ext}
\end{equation}

Both the trapezoidal and Euler evolutions are then completed by the
streaming stage, reading $\tilde{f}_{p}(\rr+h\cc_{p},t+h)=\tilde{f}_{p}(\rr,t)$
and $f_{p}(\rr+h\cc_{p},t+h)=f_{p}(\rr,t)$, respectively.

\begin{figure}
\centering{}A\includegraphics[scale=0.35]{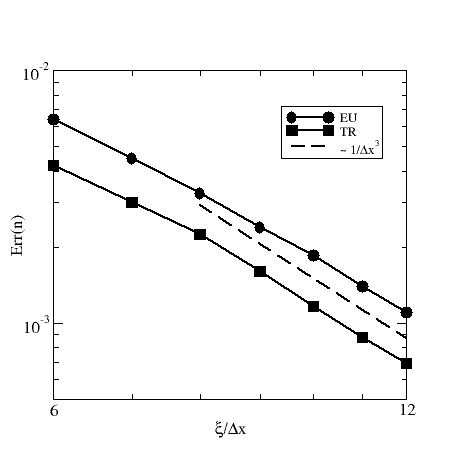}B\includegraphics[scale=0.35]{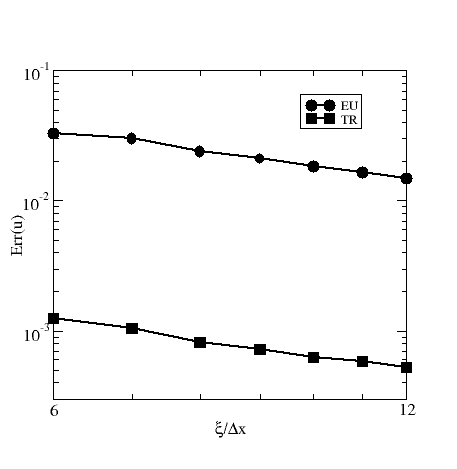}C\includegraphics[scale=0.35]{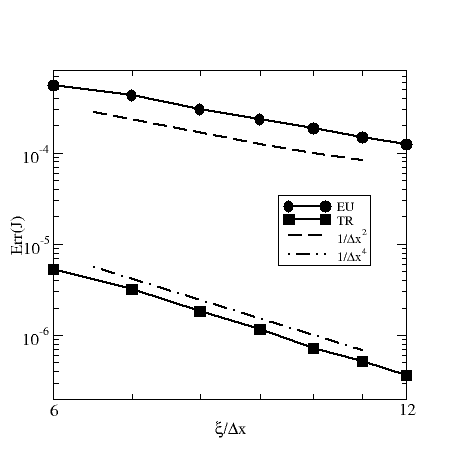}
\caption{\label{fig:NOHSerror} Numerical error in the density (panel A), velocity
(panel B) and current (panel C) for the ideal gas system in presence
of the central potential well. See text for details. The dashed and
dot-dashed lines represent the power law dependence of the numerical
error, as reported in the legends.}
\end{figure}

We simulate a three-dimensional system and in Fig. \ref{fig:NOHSerror}
we report the error on density as $Err(n)=\max_{\rr}(|n-n^{eq}|/n_{0})$,
the error on fluid velocity arising from parasitic effects, as $Err(\uu)=\max_{\rr}(|\uu|/v_{T})$,
and the error on current as $Err(\JJ)=\max_{\rr}(|\JJ|/n_{0}v_{T})$.
The data show that the numerical errors in the density, velocity
and current decrease systematically with the mesh resolution for both the EU and TR 
methodologies.
The error is reduced by about two orders of magnitude for the TR method as compared 
to the EU scheme. In particular, the error in density decreases as $\Delta x^{2}$
for both methods, while the error in current drops as $\Delta x^{2}$
and $\Delta x^{4}$ for the EU and TR methods, respectively.
These preliminary results provide a reference for the subsequent simulations of the hard-sphere
system and an important indication on the quality of the
trapezoidal evolution method.

\subsection{Hard sphere fluid mixture in a potential well}
We now consider a non ideal fluid mixture of  hard spheres, and numerically solve the statics
of the problem in the presence of the same central external potential
of eq. \eqref{eq:extpotential}, and integrate the dynamics with and
without the auxiliary distribution method. 

The trapezoidal integration for the two distributions generalizes
eq. \eqref{eq:newevolution} to 
\begin{eqnarray}
\tilde{f}_{p}^\alpha(\rr+\cc_{p}h,t+h) & = & \tilde{f}_{p}^\alpha(\rr,t)+h\, \Omega_{f,p}^\alpha(f_{p}^\alpha,\{M\})(\rr,t)
\label{eq:newevolution-two}\\
\tilde{g}_{p}^\alpha(\rr+\cc_{p}h,t+h) & = & \tilde{g}_{p}^\alpha(\rr,t)+h \, \Omega_{g,p}^\alpha(g_{p}^\alpha,\{N\})(\rr,t)
\end{eqnarray}
 where 
\begin{eqnarray}
\Omega_{f,p}^\alpha(f_{p},\{M\}) & = & \omega(f_{p}^{\alpha,eq}-f_{p}^\alpha)+S_{f,p}^\alpha
\label{eq:kernel-f}\\
\Omega_{g,p}^{\alpha}(g_{p},\{N\}) & = & \omega(g_{p}^{\alpha,eq}-g_{p}^\alpha)+S_{g,p}^\alpha
\label{eq:kernel-g}
\end{eqnarray}
 Here, $\{M\}$ and $\{N\}$ refer to the set of moments of the populations
$f_{p}^\alpha$ and $g_{p}^\alpha$ respectively.

We compute the relevant moments, that is, densities, HS chemical potentials
and currents as 
\begin{eqnarray}
\tilde{n}^\alpha & = & \sum_{p}\tilde{f}_{p}^\alpha
\label{mom-f-1-1}\\
\tilde{\Pi}^\alpha & = & \sum_{p}\tilde{g}_{p}^\alpha
\\
\tilde{\JJ}^\alpha & = & \tilde{n}\tilde{\uu}^\alpha =\sum_{p}\cc_{p}\tilde{g}_{p}^\alpha \label{mom-g-1-1}
\end{eqnarray}
 then 
\begin{eqnarray}
n^\alpha  & = & \sum_{p}f_{p}^\alpha 
\label{mom-f}\\
\Pi^\alpha  & = & \sum_{p}g_{p}^\alpha =\tilde{\Pi}^\alpha -\frac{h}{2}\CC^{\alpha ,mf}\cdot\uu^\alpha 
\\
\JJ^\alpha  & = & n^\alpha \uu^\alpha =\sum_{p}\cc_{p}g_{p}^\alpha =\tilde{\JJ^\alpha }+\frac{h}{2}\left[\CC^{\alpha ,mf}+\CC^{\alpha ,visc}+n^\alpha \FF^{\alpha }\right]
\label{mom-g}
\end{eqnarray}

The explicit form of the r.h.s. of eqs. \eqref{eq:kernel-f}-\eqref{eq:kernel-g}
reads 
\begin{eqnarray}
f_{p}^{\alpha ,eq} & = & w_{p}\left[n^\alpha+n^\alpha\uu^\alpha\cdot{\cal H}_{p}^{(1)}+n^\alpha
(2 \uu^\alpha\uu-\uu\uu):{\cal H}_{p}^{(2)}\right]
\label{eq:kernel-f-1}\\
g_{p}^{\alpha ,eq} & = & w_{p}\left[\Pi^\alpha+n^\alpha\uu^\alpha\cdot{\cal H}_{p}^{(1)}+n^\alpha
(2 \uu^\alpha\uu-\uu\uu):{\cal H}_{p}^{(2)}\right]\\
S_{f,p}^\alpha & = & w_{p}\left\{ \left({\bf C}^{\alpha,mf}+{\bf C}^{\alpha,visc}+n^\alpha\FF^{\alpha}\right)\cdot\left[{\cal H}_{p}^{(1)}+2{\cal H}_{p}^{(2)}\cdot\uu^\alpha\right]\right\} \\
S_{g,p}^\alpha& = & w_{p}\left\{ {\bf C}^{\alpha,mf}\cdot\left[\frac{\uu^\alpha}{v_{T}^{2}}+4{\cal H}_{p}^{(2)}\cdot\uu^\alpha\right]+\left({\bf C}^{\alpha,visc}+n^\alpha\FF^{\alpha}\right)\cdot\left[{\cal H}_{p}^{(1)}+2{\cal H}_{p}^{(2)}\cdot\uu^\alpha\right]\right\} 
\label{eq:kernel-g-1}
\end{eqnarray}

\begin{figure}
\centering{}
\includegraphics[scale=0.5]{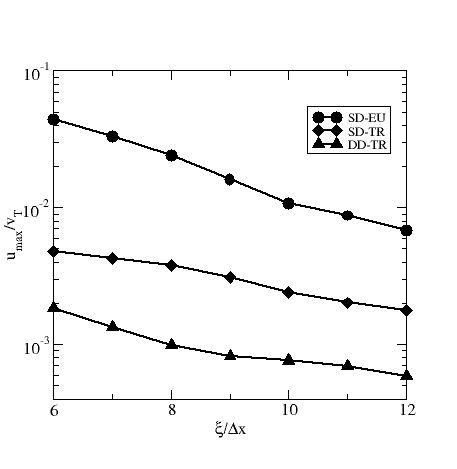}
\includegraphics[scale=0.5]{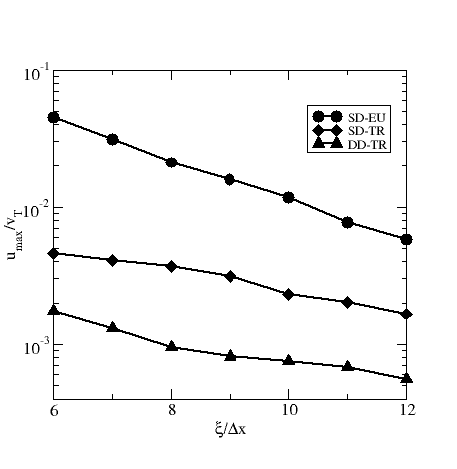}
\caption{\label{fig:HSerror}
Numerical error in the fluid
velocity in the presence of the central potential well for the one component fluid (left panel) and the
binary mixture with diameters of $\sigma_A=4$ and $\sigma_B=8$ (right panel).}
\end{figure}

In Fig. \ref{fig:HSerror}, the numerical error in the computed velocity
profiles is reported for the one component and the two component fluids. 
The error decreases with increasing resolution
and is smaller by respectively a factor $10$ and $50$ for the case
of SD-TR and DD-TR simulations as compared to the SD-EU method. The
data are similar for the one and two-component systems,
follow the same behavior observed for the ideal gas, and the
error in velocity decreases steadily with increasing resolution. The
spurious velocities are about 50 times smaller for the DD-TR case
as compared to the SD-EU integration.

A major advantage of the trapezoidal integration alone is the possibility
to work at high packing fractions, up to about $0.35$, whereas with
standard Euler integration, the maximal packing fraction before numerical
instabilities develop is $0.27$.


\subsection{Channel flow with a bottleneck}

\begin{figure}
\begin{centering}
\includegraphics[scale=0.35]{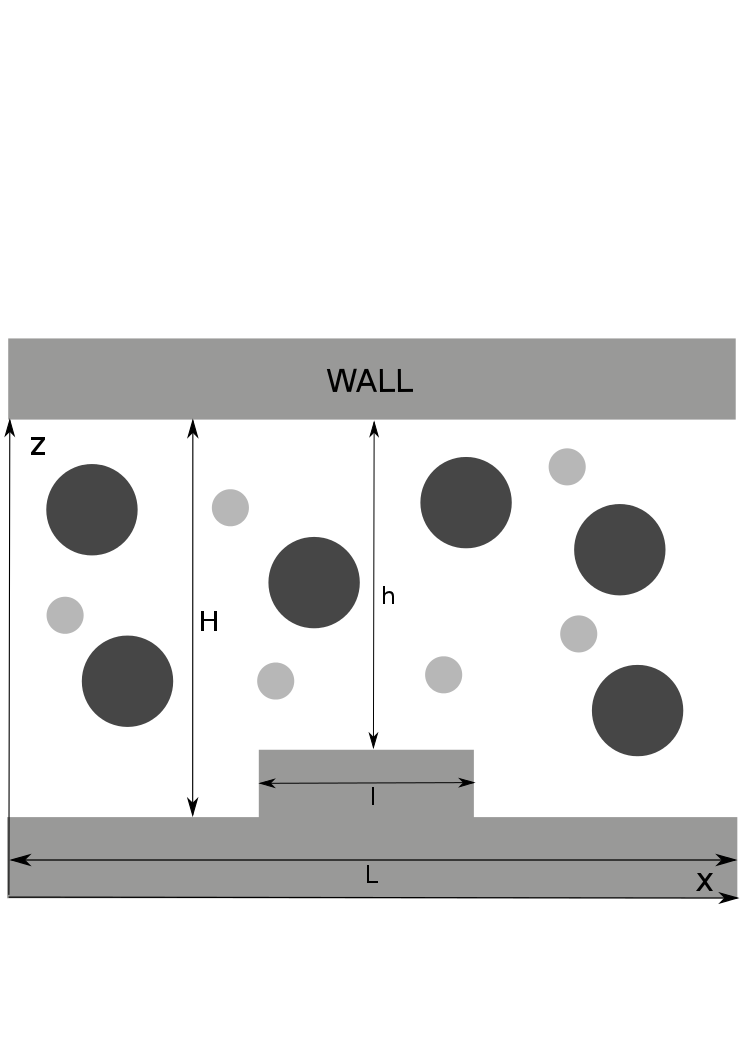}
\par\end{centering}

\centering{}
\caption{\label{fig:sketch-obstacle}Sketch of the channel flow system in presence
of an obstacle. The channel is filled with either a one-component
fluid or a binary mixture. For the simulations we have set the geometry
to $L=120$, $l=30$, $H=35$ and $h=28$ in lattice units. The diameters of the large and small
particles are denoted $\sigma_A$ and $\sigma_B$, respectively.}
\end{figure}

We now consider the flow of a one-component HS fluid and a binary
mixture in a channel flow, in the presence of a bottleneck, as
depicted in Fig. \ref{fig:sketch-obstacle}. Flows in the presence of
a sharp obstacle represent a critical test to the numerical methodology
due to the harsh collisions that the particles experience with the corners of the
obstacle. In particular, we choose a rather strong forcing term, being
equal to $10^{-3}$ in lattice units, in order to obtain large impinging
velocities against the obstacle. We further impose no-slip boundary
conditions on the fluid populations at the solid wall for both the $f$ and the $g$
distributions. For this we employ the mid-point bounce-back rule on the populations \cite{LBgeneral}.  
{We initially simulate a one-component system composed of hard spheres 
of diameter $\sigma=8$ and make complementary simulations with a two-component mixture 
with hard spheres diameters of $\sigma_A=4$ and $\sigma_B=8$.}

\begin{figure}
\begin{centering}
\includegraphics[scale=0.4]{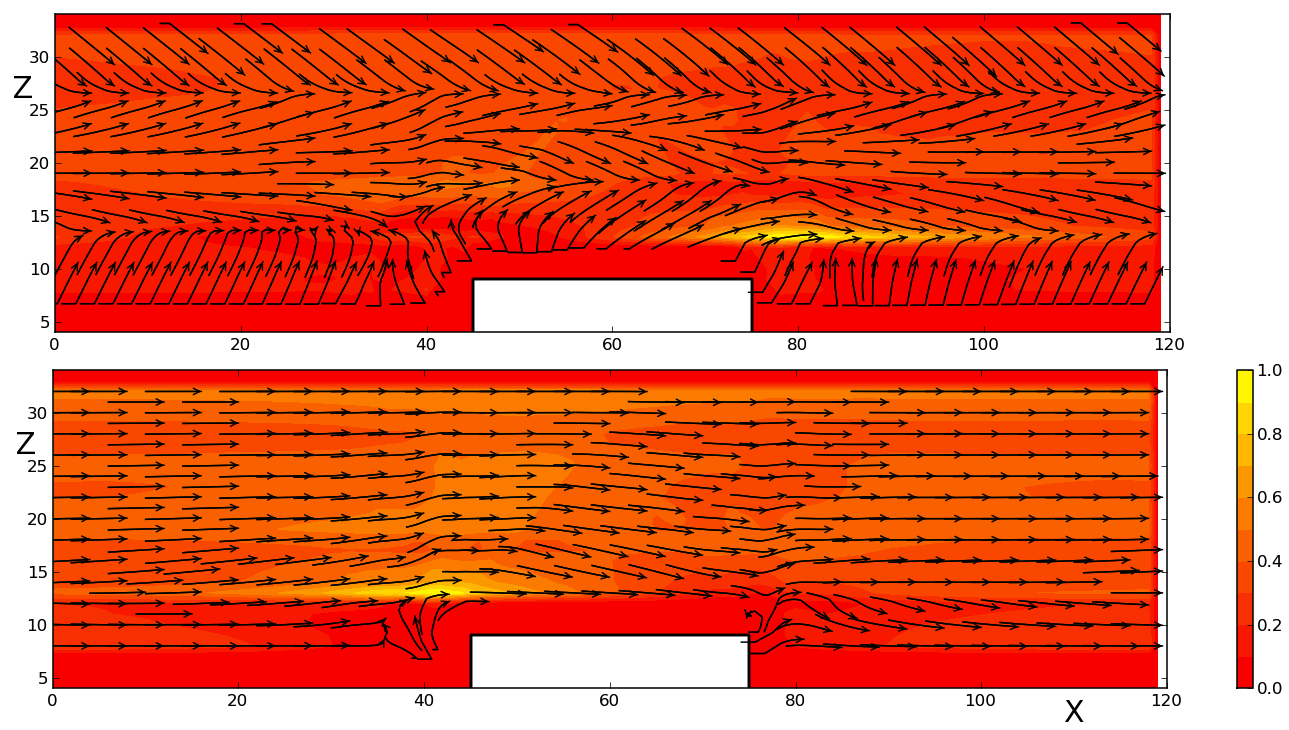}
\par\end{centering}

\centering{}
\caption{\label{fig:striction-comparison} (Color online) One component system: comparison
of streamlines for packing fractions of 0.13 as obtained with the
SD-EU (upper panel) and with the DD-TR (lower panel) methods.}
\end{figure}

As Fig. \ref{fig:striction-comparison} demonstrates, the naive SD-EU
method provides strong spurious velocities arising from the presence
of the wall. In fact, away from the obstacle, the streamlines are
expected to be parallel to the wall, whereas we observe strong non-parallel
streamlines near the wall that confirm the low quality of this type of simulations.
Conversely, the DD-TR method provides well aligned streamlines near
the wall and far away from the bottleneck.
From these observations, we decided to consider further benchmarks by looking 
at the results obtained with the DD-TR methodology alone.


\begin{figure}
\begin{centering}
\includegraphics[scale=0.5]{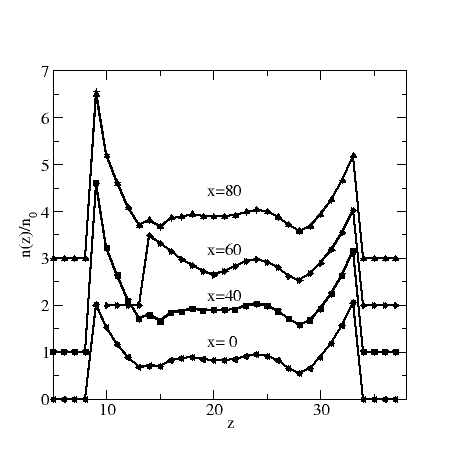}
\par\end{centering}

\centering{}
\caption{\label{fig:densityprofiles}One component system: density profiles
for different values of the x coordinate as obtained with and without
forcing. The profiles for $x=40$ and $x=80$ correspond to positions right before
and right after the corners of the obstacle. For all $x$ values, the
profiles with and without forcing are basically indistinguishable.
Profiles have been shifted upward for the sake of clarity.}
\end{figure}

\begin{figure}
\begin{centering}
\includegraphics[scale=0.5]{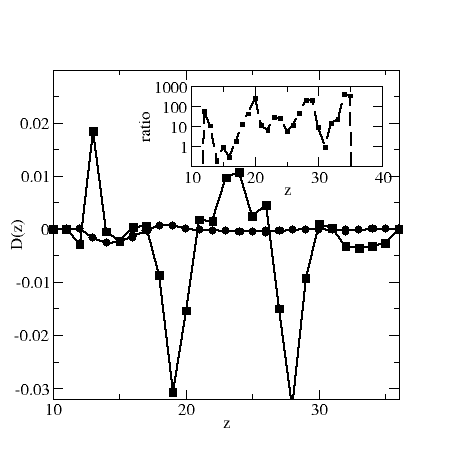}
\par\end{centering}
\centering{}
\caption{\label{fig:divvelprofiles}One component system: divergence of velocity
$D \equiv \nabla \cdot \uu / (v_T / \Delta x)$ computed at mid channel ($x=60$) for the system in flow condition. 
The two profiles correspond to the DD-TR (circle symbols) and SD-EU (square symbols) methods. 
The inset  displays the  ratio $|D^{DD-TR} / D^{SD-EU}|$.
}
\end{figure}

{Hard spheres in proximity to an irregular surface present an interesting phenomenology
in itself.  In fact,}
in proximity to the obstacle, the fluid particles go around the obstacle
with non-trivial patterns. In particular, as the flow lines 
in Fig. \ref{fig:densityprofiles} reveal, a first bounce back is found
near the convex corner. Entropic forces have a strong influence on
{the spatial distribution of the particles and, as previous studies demonstrate
\cite{Dietrich}, the concave corners effectively attract particles, while convex
corners exert repulsive forces}. Such  dual behavior is recovered by our
simulations, as revealed by the density profiles in Fig. \ref{fig:densityprofiles}
where the accumulation of particles toward the edges of the obstacle
is clearly visible. In addition, we observe that the density profiles
have a very weak dependence on the forcing term, with a somehow stronger
variation in proximity of the corners for the incoming particles, as compared to the static case. 
A further validation of the method is given by the computation of the divergence of velocity,
as reported in Fig. \ref{fig:divvelprofiles}. For the compressible system considered here, the 
quantity $\nabla \cdot \uu$ should be zero everywhere, while spurious compressibility effects are clearly visible 
when employing the SD-EU method. The DD-TR method minimizes such error up to three order of magnitudes.

For the system at hand, the simulations
provide new interesting information about the fluid velocity in this
geometry, as shown in the following.

\begin{figure}
\begin{centering}
\includegraphics[scale=0.3]{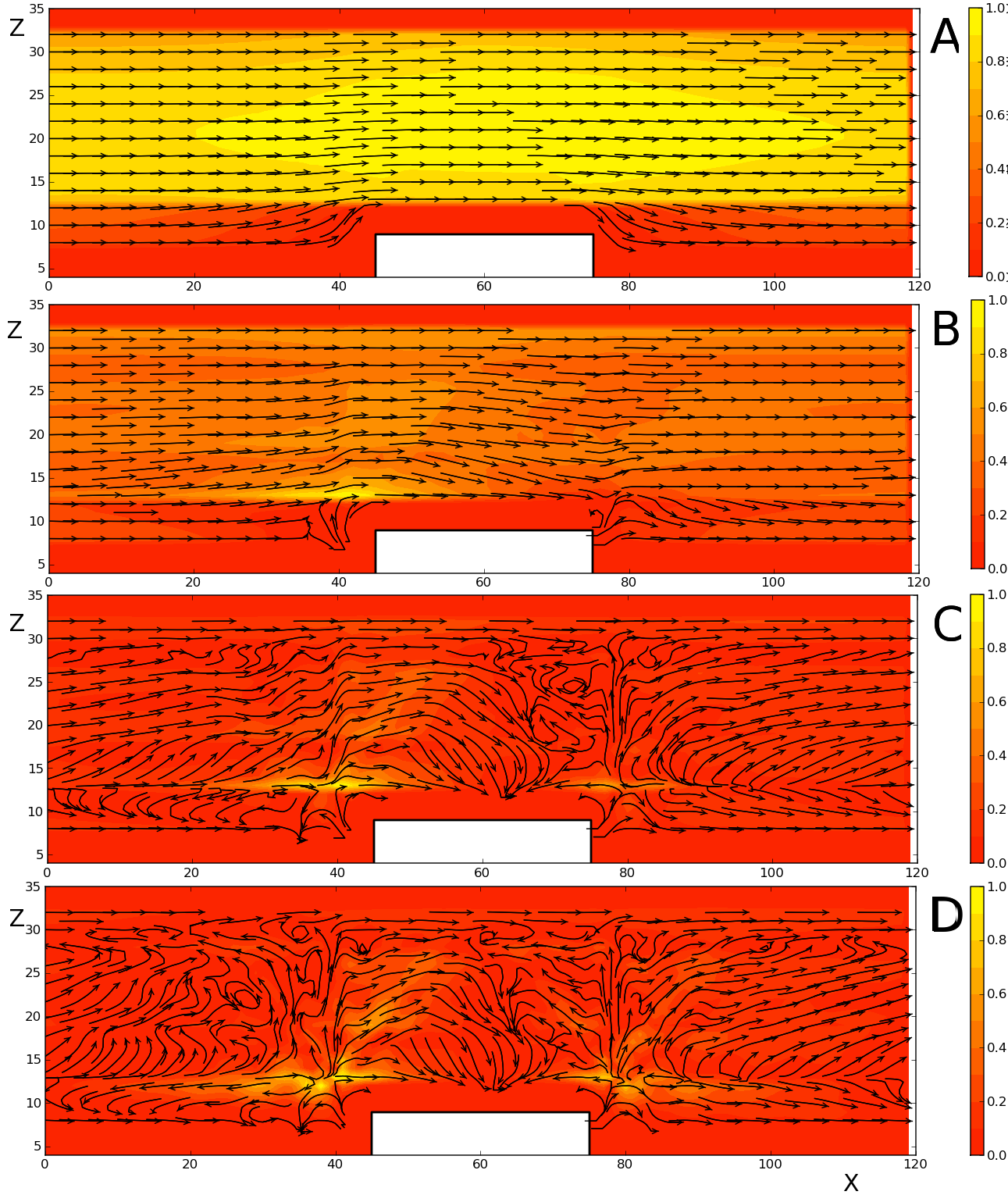}
\par\end{centering}
\centering{}
\caption{\label{fig:strictiononecomponent} (Color online) One component system: streamlines
for packing fractions of 0 (A panel), 0.13 (B panel), 0.26 (C panel)
and 0.34 (D panel). The color map represents the modulus of the flow
velocity, normalized by its maximum.}
\end{figure}

The simulations provide the fine details of
flow pattern for the one-component system at varying packing fraction,
as illustrated in Fig. \ref{fig:strictiononecomponent}. For increasing
packing fraction, the streamlines become more and more disordered
in proximity to the convex corners of the bottleneck. A quite disordered pattern is
observed already at a packing fraction of $0.26$, with flow
separation appearing in correspondence with the impinged corner. The
dynamical disorder appears to initiate at the far away edge of the obstacle
with respect to the incoming flow direction. At a packing fraction of
$0.34$, the disorder has propagated to the whole region of the bottleneck
with rough recirculation patterns. It should be noticed that, for increasing
packing fraction, the modulus of velocity is reduced overall, with
strong peaks localized near the corners. 

\begin{figure}
\begin{centering}
\includegraphics[scale=0.2]{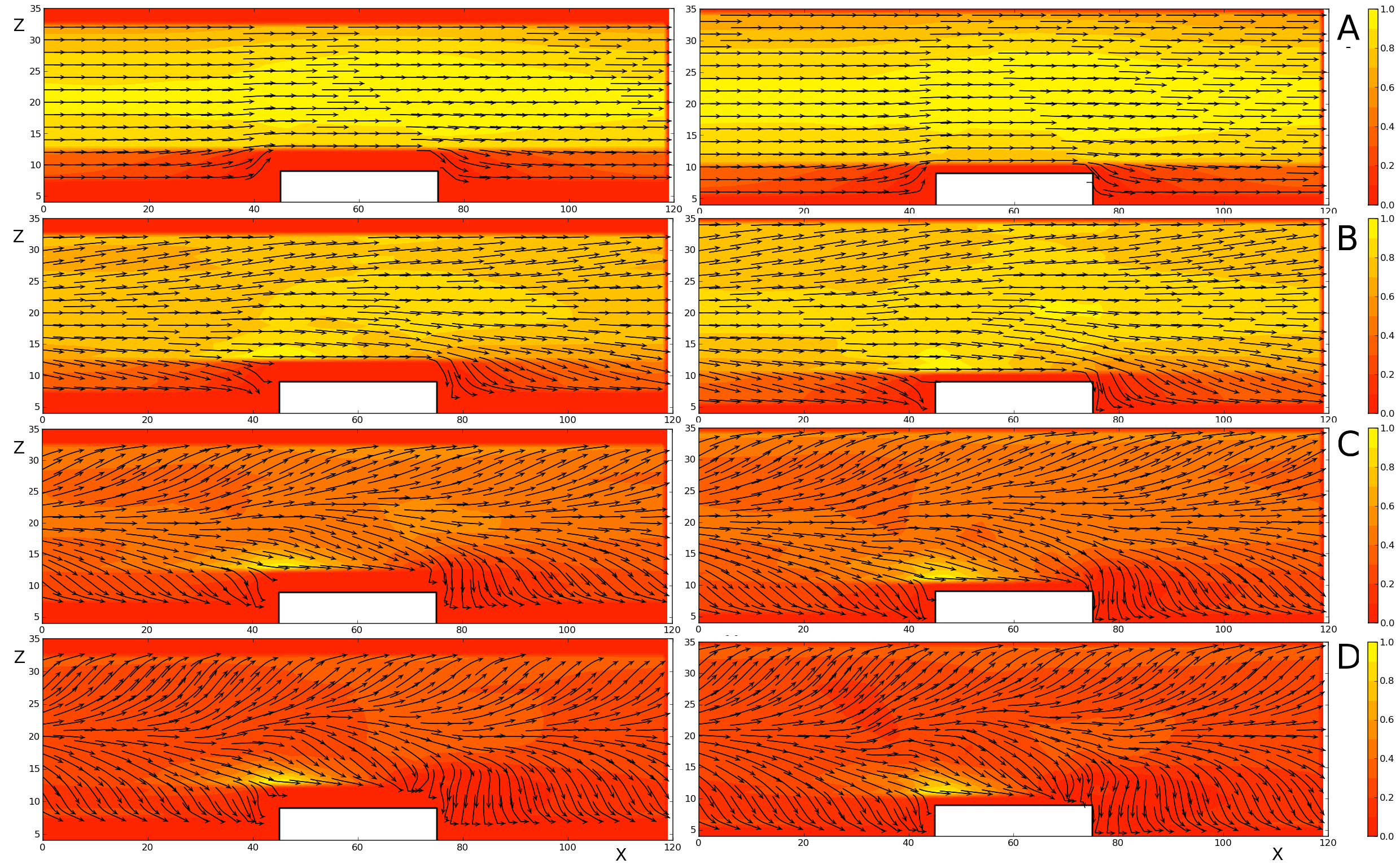}
%
%
%
\par\end{centering}
\caption{\label{fig:strictionmixture} (Color online) Binary mixture with $\sigma_{A}=8$ (left
column) and $\sigma_{B}=4$ (right column): streamlines for packing
fractions of 0.0 (panel A), 0.13 (panel B), 0.26 (panel C) and 0.34
(panel D) and for 50\% composition. }
\end{figure}

\begin{figure}
\begin{centering}
\includegraphics[scale=0.21]{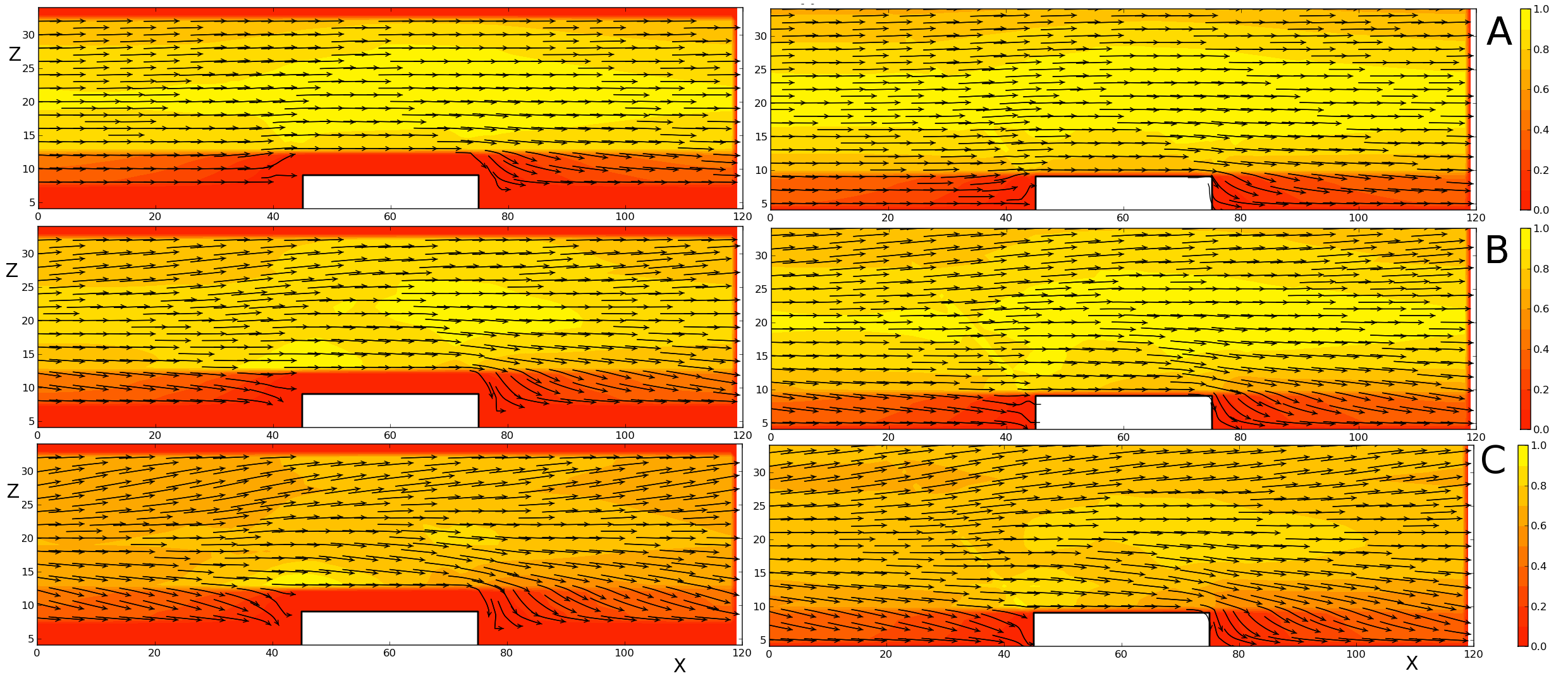}
%
%
\par\end{centering}

\caption{\label{fig:strictionmixture8:2} (Color online) Binary mixture with $\sigma_{A}=8$
(left column) and $\sigma_{B}=2$ (right column): streamlines for
packing fractions of 0.13 (panel A), 0.26 (panel B) and 0.34 (panel
C) and for 50\% composition.}
\end{figure}

We have next considered a binary mixture of hard spheres of diameters $\sigma_A$ and $\sigma_B$, 
flowing in the same channel
with the bottleneck. 
An important aspect of the binary mixture is that entropic
forces play different roles on the species with different diameters.
For particles of smaller diameter, entropic forces are smaller, and these particles
can distribute more uniformly between the concave and convex corners.
Consequently, the flow pattern is expected to be more ordered. This
behavior is shown in Fig. \ref{fig:strictionmixture}, for a binary mixture
with $\sigma_{A}=8$ and $\sigma_{B}=4$. The streamlines of both
the large and small particles have a smoother behavior as compared
to the one-component case. The modulus of velocity of both species
is more uniform as compared to the one-component system, with a smoother
distribution around the obstacle. In Fig. \ref{fig:strictionmixture8:2},
the binary mixture with particles of size $\sigma_{A}=8$ and $\sigma_{B}=2$ presents even
smoother flow lines and smoother distribution of the velocity moduli, as compared to simulations 
at smaller size ratio and at the corresponding packing fractions. 
Overall, we conclude that in the binary mixture,
the component with particles of smaller size acts as a powerful lubricant that regularizes
the flow pattern and distributes evenly the flow velocity over the whole system.

\section{Conclusions}
\label{conclusions}
In this paper, we have illustrated a numerical version of the Lattice Boltzmann method 
for the simulation of hard-sphere one-component and binary mixtures,
that can deal with rapid spatial variations in the number density. 
As well-recognized in the Lattice Boltzmann community, strong inhomogeneities in the density 
induce strong parasitic currents that need to be handled with great care. 

In our method, we have extended the previous ideas of He et
al. \cite{He} 
and T. Lee \cite{Lee}
but with some important modifications. In particular, we compute the excess chemical potential 
arising from the hard sphere collisions on-the-fly, without resorting to an educated guess 
of its functional form.  In addition, we have adapted the trapezoidal integration rule for the time evolution
of the populations, written as an explicit time-stepping algorithm. 

The numerical results showed that, at all packing fractions considered in the benchmarks,
the method provides robust results and stable numerical behavior.

We conclude by mentioning that the presented method can be applied without major modifications 
to nanofluids in presence of electrostatic interactions, as presented in ref. \cite{NOIEPL}. 
For these systems, internal electrostatic forces exerted between charged species arise 
from the solution of a Poisson problem treated at mean-field, Vlasov level. Also in this case, 
the trapezoidal and double distribution methodology can be applied straightforwardly since 
electrostatic forces are treated at the same level of external forces.

\section{Appendix}
We report the formulae given elsewhere which have been used to compute the
various terms of the effective field. 
The details have been reported in a previous publication \cite{JCP2011}.
In eq. \eqref{splitforce}  we can identify a force acting on the $\alpha$ particle at $\rr$ due to
the influence of all remaining particles in the system, the so called potential of mean force. For a hard-sphere mixture we have:
\be
\CC^{\alpha,mf}(\rr,t)=-k_B T\na(\rr,t)\sum_\beta\sigma_{\alpha\beta}^2 
\int d\bk \bk
g_{\alpha\beta}(\rr,\rr+\sab\bk,t)
n_{\beta}(\rr+\sab\bk,t)+\na(\rr,t)\sum_\beta \GG^{\alpha\beta}(\rr,t)
\ee
with $\sigma_{\alpha\beta}=(\sigma_{\alpha\alpha} +\sigma_{\beta\beta})/2$, while the last term represents  the molecular fields associated with the attractive forces:
\be
\GG^{\alpha\beta}(\rr,t)=
- \int dr' \nb(\rr',t)\gab(\rr,\rr')\nabla U^{\alpha\beta}(\rr-\rr')
\ee
with $U^{\alpha\beta}(r)$ a long range attractive potential.
The drag term is:
\be
\CC^{\alpha,drag}(\rr,t) \simeq
-\na(\rr,t)\sum_\beta 2\sigma_{\alpha\beta}^2 \sqrt{\frac{ k_B T}{\pi} }
\frac{4\pi}{3}
g_{\alpha\beta}(\{ \na(\rr,t)\})
\nb(\rr,t) (\uua(\rr,t)-\uub(\rr,t)) 
\label{dragforce2}
\ee
and for the viscous part
\be
\CC^{\alpha,visc}(\rr,t)=\na(\rr,t)
\sum_\beta 2  \sigma_{\alpha\beta}^2 \sqrt{\frac{m k_B T}{\pi} }
\int d\bk \bk
g_{\alpha\beta}(\rr,\rr+\sab\bk,t)
\nb(\rr+\sab\bk,t)
 \bk\cdot
(\uub(\rr+\sab\bk)-\uub(\rr)) 
\label{viscousforce}
\ee
where $g_{\alpha\beta} $ is the pair correlation function evaluated  at contact ($r=\sigma_{\alpha\beta}$)  
As shown in Ref. \cite{JCP2011} one can derive the following expressions in the limit of a uniform system for the viscosity:
\be
\eta^{\alpha\beta}=\frac{4 \pi}{15}
\sigma_{\alpha\beta}^4 \sqrt{\frac{m k_B T}{\pi} }g_{\alpha\beta} \nb
\ee
and
\be
\eta_b^{\alpha\beta}=\frac{5}{3}\eta^{\alpha\beta}
\ee


\subsection{Acknowledgments}

We thank Jonas L\"att for drawing to our attention to ref. \cite{Lee}
and Benjamin Rotenberg for suggesting the use of the trapezoidal rule.
 
 \end{document}